

\magnification=1200
\baselineskip=14pt

\def\cl{\centerline}

\ \

\bigskip\bigskip
\bigskip\bigskip
\bigskip\bigskip

\cl{\bf GAMMA RAY BURSTS ARE TIME-ASYMMETRIC }

\bigskip\bigskip
\bigskip\bigskip

\cl{ R.J. Nemiroff$^{(1),(2),(3)}$, J. P. Norris$^{(3)}$, }
\cl{ C. Kouveliotou$^{(1),(4)}$, G.J. Fishman$^{(4)}$, }
\cl{ C.A. Meegan$^{(4)}$, and W.S. Paciesas$^{(4),(5)}$ }

\bigskip\bigskip
\bigskip\bigskip

\cl{\it $^{(1)}$ Universities Space Research Association }
\cl{\it $^{(2)}$ George Mason University, Fairfax, VA 22030 }
\cl{\it $^{(3)}$ NASA Goddard Space Flight Center, Greenbelt, MD 20771 }
\cl{\it $^{(4)}$ NASA Marshall Space Flight Center, Huntsville, AL 35812 }
\cl{\it $^{(5)}$ University of Alabama, Huntsville, AL 35899 }

\bigskip\bigskip
\bigskip\bigskip
\bigskip\bigskip
\bigskip\bigskip
\bigskip\bigskip
\bigskip\bigskip

\cl{ In press: The Astrophysical Journal }

\vfill\eject
\baselineskip=18pt

\ \
\bigskip\bigskip

\cl{\bf ABSTRACT }

\bigskip

A simple test for time-asymmetry is devised and carried out on the
brightest gamma-ray bursts detected by the Burst and Transient Source
Experiment (BATSE) on board the {\it Compton} Gamma Ray Observatory. We
show evidence that individual bursts are time-asymmetric on all time scales
tested, from a time scale shorter than that of pulses which compose GRBs to
a time scale similar to a greater envelope that contains these pulses. We
also find bursts which manifest significant asymmetry only on time scales
comparable to the duration of burst, and bursts for which no clear
asymmetry on any time scale is present. The sense of the asymmetry is that
bursts and/or component structures rise in a shorter time than they decay.
We also find that our whole sample of bursts taken together is
time-asymmetric, in that there are significantly more bursts and pulses
where the rise is more rapid than the decay, on all time scales tested and
for all energy bands tested. When our whole GRB sample is binned at 64-ms
and integrated over all BATSE energies, the statistical significance is at
the 6 $\sigma$ level. Models that predict time-symmetry are therefore
excluded.

\bigskip

\noindent {\it Subject headings:} gamma-rays: bursts

\vfill\eject

\cl{\bf 1. INTRODUCTION }

\bigskip

There is no present consensus as to the cause or location of gamma-ray
bursts (GRBs).  GRBs were first discovered in data from the Vela satellites
as reported by Klebesadel, Strong, and Olson (1973), and since have been
detected by numerous spacecraft. Recent results by the Burst and Transient
Source Experiment (BATSE) on board the {\it Compton} Gamma Ray Observatory
have confirmed their isotropic angular distribution and ``confined" peak
brightness distribution (Meegan et al. 1992). These results generate an
interesting puzzle.  Currently, there are more than 100 models for the
origins of GRBs (Nemiroff 1993).  To help eliminate inapplicable models,
discriminating statistical tests are needed.

Some GRB models depend on processes or geometries that are time-symmetric.
A certain amount of relativistic motion is essential to many GRB models to
avoid being dominated by collisions between gamma-rays.  Were these models
to rely on a relativistic beam crossing our line of sight in order to
generate the time structure inherent in GRBs, this reliance would make the
resulting light curves, when taken together, time-symmetric (Link, Epstein
and Priedhorsky 1993, hereafter LEP).  This is because there is an equal
chance that the beam sweeps past us going in one direction as there is in
the other direction.

At least one specific model (McBreen \& Metcalfe 1988) relies on a
gravitational lensing scenario to generate GRB light curves.  This would
also produce a distribution of light curves that, when taken together,
would be time-symmetric, as there is an equal chance of the source crossing
gravitational lens caustics going in either direction.

Time-asymmetry in GRB light curves has been discussed by several authors
(see Barat et al. 1984; Norris et al. 1987; Mitrofanov et al. 1992,
Kouveliotou et al. 1992; Fishman et al. 1993; and by LEP).  Mitrofanov et
al. (1992) computed an asymmetry in the durations and hardness ratios
between the rising parts of bright GRB light curves and the falling parts.
LEP used a {\it skewness function} which measured temporal asymmetry for a
majority of 20 bright GRBs taken from the first 48 detected by BATSE. LEP's
skewness function is similar to that used by Weisskopf et al. (1978) on
Cygnus X-1 data.

In this paper we compute a simple, easily measured parameter designed to
quantify the degree of time-asymmetry present in GRB time profiles. This
parameter is compared to that expected with a time-symmetric GRB light
curve. Many GRB light curves are seen to be composed of a series of
individual pulses (Norris et al. 1993, Davis et al. 1993). Figure 6 of
Barat et al. (1984) have reported that short, single-peaked GRBs
detected by the Franco-Soviet instrument Signe on board {\it Venera}
spacecraft have a rise-to-decay time ratio of about 0.321. Norris et al.
(1993) indicates that individual isolated pulses tend to have a
rise-to-decay time ratio of about 0.67, although a wide range is observed.

Our asymmetry test differs from that in LEP in that its statistical
significance can be easily computed.  Also, our test involves a cumulative
sum which is taken over a complete sample of GRBs, allowing us to see if an
asymmetry accumulates over many GRBs.  Our asymmetry test differs from
Mitrofanov et al. (1992) in that our test can be easily conducted over a
wide range of time scales. In \S 2 we describe the time-symmetry parameter
we use and the attributes of the data set tested. In \S 3 we present the
results of the asymmetry analysis.  We discuss these results and inferences
in \S 4.

\bigskip

\cl{\bf 2. STATISTICS AND DATA }

\bigskip

The time-asymmetry parameter we use is the ratio of the number of times
where the counts in the previous time bin are higher to the number where
the counts are lower - a statistic we call $\Upsilon$. The number counts
that $\Upsilon$ is based on begin with the first recorded bin of the GRB
time series that is 9 $\sigma$ above a fit background level, and continue to a
time where the GRB has subjectively been deemed to have concluded. At this
time a background fit begins and typically runs for 50 seconds.

Each sample in the time series is inspected to see whether it is 9 $\sigma$
above a predetermined background level. If it is, the counts in this bin
are compared to the counts in the previous bin.  If the counts in the bin
under consideration are greater (less) than the counts in the preceeding
bin, the number of rising (falling) bins is incremented by one. The bin
succeeding each 9 $\sigma$ or greater sample bin is also tested, regardless
of whether it is 9 $\sigma$ above the background or not.  If the number
counts in this bin are greater (less) than the counts in the preceeding
bin, the number sum for the number of rising (falling) bins is increased by
one. As stated above, $\Upsilon$ is the ratio of these two number sums.
Demanding that the next bin after every 9 $\sigma$ bin be tested is done to
insure that the test itself does not introduce a time-asymmetry. The
minimum bin value of 9 $\sigma$ was used to insure that each bin by itself
was of sufficient magnitude to meet BATSE's trigger criterion. Therefore,
we virtually eliminate biases which favor quickly rising GRBs over slowly
rising GRBs, within the constraint that each GRB must have triggered the
BATSE instrument.

The $\Upsilon$ statistic is particularly easy to compute and test for
significant deviation from symmetry, for which $\Upsilon$ is unity. A
time-symmetric profile would have an equal number of rising and falling
bins.  The statistics of $\Upsilon$ is a direct extension of the statistics
of flipping a coin, which is well understood in terms of binomial and
normal distributions.

We use data with 64-ms time resolution (BATSE PREB + DISCSC data types,
Fishman et al. 1989) to compute $\Upsilon$ for a wide range of time scales:
from 64 ms to 4096 ms. 64 ms is the shortest accumulation trigger time
scale employed by the BATSE instrument.  Above the upper limit of 4096 ms,
the statistical significance of the test begins to drop.

We have applied this test to a sample of the brightest GRBs detected by
BATSE up to 1993 March 10. Specifically, we utilize bursts with peak count
rates greater than 18000 counts sec$^{-1}$, and durations greater than 1
second. These brighter GRBs are easy to distinguish and give good
statistics.

\bigskip

\cl{\bf 3. RESULTS }

\bigskip

We were able to confirm with clear statistical significance the generally
believed result (see, for example, Barat 1984, Fishman et al. 1993, and
LEP) that GRBs light curves are time-asymmetric. For the sample discussed
in \S 2, the $\Upsilon$ statistic is shown in Figure 1, for 64-ms time
bins, and for the energy window from 50 to 300 keV. A GRB that is
time-symmetric would have a ratio of unity, and a sample of time-symmetric
GRBs would be evenly distributed about unity. From inspection of Figure 1
it is clear that most GRBs have more falling bins than rising ones.

The error bars depict one $\sigma$ standard deviation in the number of bins
counted to be rising or falling, assuming this number is normally
distributed about $1/2$ the total number of bins counted: $n_{t}/2$. More
precisely, the total number of time bins is the sum of rising and falling
bins: $n_{t}= n_{r} + n_{f}$. The values of $\Upsilon$ at the one $\sigma$
upper and lower bound error bar are then $(n_{r} + \sqrt{n_{t}/4})/(n_{f} -
\sqrt{n_{t}/4})$ and $(n_{r} - \sqrt{n_{t}/4}) / (n_{f} + \sqrt{n_{t}/4})$,
respectively.

Some GRBs are seen to have a statistically significant excess of falling
bins over rising bins, while others do not. None of the GRBs yielded
$\Upsilon$ significantly greater than unity. It is straightforward to
compute a cumulative $\Upsilon$ statistic for all the GRBs in the sample,
to see if the asymmetry averages out over many GRBs, or accumulates to
become more significant.  To compute a cumulative $\Upsilon$, we counted
the total number of rising and falling bins for all the GRBs in the whole
sample, and then computed the ratio. The resultant value for $\Upsilon$ on
the 64-ms time scale and integrated over the complete BATSE energy range
(above 25 keV) is 0.898 ${+ 0.018 \over - 0.016}$. Thus, for this
group of GRBs, the time-asymmetry result is significant at the 6 $\sigma$
level, which corresponds to better than one part in $10^{-8}$.

Figure 2 shows $\Upsilon$ for three individual GRBs and for the complete
sample for the energy band 50 - 300 keV.  The light curves for these GRBs
are shown in Figures 3a, 3b, and 3c.  The size of the bins was varied from
64 ms to 4096 ms.  The points plotted in Figure 2 were artificially
staggered in time (slightly) so that the error bars would be clearly
visible for each.

As seen in Figure 2, BATSE trigger 143 shows significantly more falling
bins than rising bins on all time scales.  The asymmetry of this GRB is
clearly evident in Figure 3a, from the asymmetric shape of the first
envelope of gamma-ray pulses. A general trend is evident in this burst:
pulse intensity decreases as the burst progresses. This behavior drives the
asymmetry on the 1.024, 2.048 and 4.096 second time scales. On shorter
time scales, the asymmetry is still evident because the individual pulses
themselves are time-asymmetric, a result previously reported by Norris et
al. (1993). In contrast, LEP reported asymmetry only on longer time scales.

Next we inspect trigger 1606.  From inspection of $\Upsilon$ and this GRB's
light curve in Fig. 3c, we find no significant evidence for time-asymmetry
in this GRB, either on time scales characteristic of pulses or in the
general shape of the burst profiles or envelopes.

Finally, we inspect trigger 678.  This GRB is especially interesting since
the general envelope of pulses is significantly asymmetric (evident from a
visual inspection of Fig. 3b), while there is no strong evidence from the
asymmetry parameter that the individual pulses themselves are asymmetric.
This is shown in Fig. 2 by the lack of a significant deviation of
$\Upsilon$ from unity on shorter time scales, but the onset of significant
asymmetry at the longest time scale, 4096 ms.  A separate 16-ms time scale
was run on MER data for this burst, showing only a slight asymmetry with an
$\Upsilon$ not significantly different from the 64-ms result shown.

The GRB population tested as a whole is significantly time-asymmetric on
time scales from 64 ms to 4096 ms, as Fig. 2 also shows.  The error bars
denote the one standard deviation errors for the cumulative number counts
for rising and falling bins.

Figure 4 shows $\Upsilon$ as a function of bin size for the four energy
bands of the BATSE Large Area Detectors.  Inspection of this figure shows
that the GRB sample is significantly time-asymmetric not only on all
time scales tested but also in all four energy bands, and that the magnitude of
asymmetry is, within errors, independent of energy band.

\bigskip

\cl{\bf 4. DISCUSSION AND CONCLUSIONS }

\bigskip

We have verified that a group of GRBs selected so as to give good
statistics for an asymmetry test do indeed, when taken together, show a
statistically significant amount of time-asymmetry, as reflected in the
$\Upsilon$ statistic, in their light curves from 0.064 seconds to 4.096
seconds, and in four contiguous energy bands above $\sim$ 25 keV.

Additionally, we have found individual GRBs light curves which show
asymmetry in the $\Upsilon$ statistic on all time scales, on long time
scales only, and on no time scales at all. The last case is particularly
interesting, as it shows that GRBs can be asymmetric on time scales longer
than their component pulses.

Could our result be primarily due to systematic errors? One way a false
time-asymmetry could have entered the data is through an artifact of the
BATSE trigger criterion. A slowly rising very smooth GRB, which later
rapidly declines giving the GRB a marked asymmetry of $\Upsilon > 1$, might
not trigger the BATSE detectors because it would have been considered part
of the background.  Such a GRB would only show time-asymmetry, however, on
time scales comparable to the BATSE background determination time scale,
which is about 17 seconds (Fishman et al. 1992). Since such a time scale is
longer than considered here, it does not effect the analysis for the bursts
considered here individually. Taken as a group, the systematic exclusion of
slowly rising GRBs might, however, confound the result that, as a whole,
GRBs are time-asymmetric.  We do not feel this is the case as GRBs that are
smooth over the time scale of seconds are quite unusual.

We note that the GRBs that populate the short duration hump of the duration
histogram (Kouveliotou et al. 1993) have been excluded from our sample. The
existence of a separate short duration grouping of GRBs has been discussed
previously by Cline and Desai (1974), Mazets (1981), Norris et al. (1984),
Klebesadel (1990), and Hurley (1991). Realistically, we would not expect
the excluded group population to affect significantly the above quoted
results, not because of any property they might possess (see Bhat et al.
1993, and Kouveliotou et al. 1993), but rather because their short
durations would only slightly perturb the statistics.

\bigskip

We thank Jerry Bonnell for discussions and comments.  This work was
supported in part by NASA under the {\it Compton} Gamma-Ray Observatory
Guest Investigator Program.

\vfill\eject

\cl{\bf REFERENCES }
{
\parindent=0pt
\hangindent=20pt
\baselineskip=10pt
\parskip=4pt

\bigskip

\hangindent=20pt
Barat C., Hayles, R. I., Hurley, K., Niel, M., Vedrenne, G., Estulin, I.
V., and Zenchenko, V. M. 1984, ApJ, 285, 791

\hangindent=20pt
Bhat P. N., Fishman, G. J., Meegan, C. A., Wilson, R. B., and Paciesas, W.
S., 1993, in Compton Gamma-Ray Observatory, AIP Conference Proceedings 280,
eds: M. Friedlander, N. Gehrels, and D. J. Macomb (New York: AIP), 953

Cline, T. and Desai, U. 1974, Proc. 9th ESLAB Symp., 37

\hangindent=20pt
Davis, S. P. and Norris, J. P. 1993, in Compton Gamma-Ray Observatory, AIP
Conference Proceedings 280, eds: M. Friedlander, N. Gehrels, and D. J.
Macomb (New York: AIP), 964

\hangindent=20pt
Fishman, G. J. et al. 1989, in the Gamma Ray Observatory Science Workshop,
ed: W. Neil Johnson, 10-12 April 1989, Greenbelt, MD, p 2-39

\hangindent=20pt
Fishman, G. J. 1993, presentation at the {\it Compton} Observatory
Symp., Washington University, Oct. 15, in press

\hangindent=20pt
Fishman, G. J., Meegan, C. A., Wilson, R. B., Paciesas, W. S., Pendleton,
G. N. 1992, in The Compton Observatory Science Workshop, eds: C. R.
Shrader, N. Gehrels, and B. Dennis, NASA Conference Publication 3137, p 26

\hangindent=20pt
Hurley, K. et al. 1991 in Gamma-Ray Bursts, AIP Conference Proceedings 265,
eds: W. S. Paciesas and G. J. Fishman, (New York: AIP), 195

\hangindent=20pt
Klebesadel, R. W. 1990 in Gamma Ray Bursts, eds: C. Ho, R. I. Epstein, and
E. E. Fenimore, (Cambridge: Cambridge U. Press), 161

Klebesadel, R., Strong, I. B., Olson, R. A. 1973, ApJ, 182, L85

\hangindent=20pt
Kouveliotou, C., Paciesas, W. S., Fishman, G. J., Meegan, C. A., and
Wilson, R. B. 1992 in The Compton Observatory Science Workshop,
eds: C. R. Shrader, N. Gehrels, and B. Dennis, NASA Conference Publication
3137, p. 61

\hangindent=20pt
Kouveliotou C., Meegan, C. A., Fishman, G. J., Bhat, N. P., Briggs, M. S.,
Koshut, T. M., Paciesas, W. S., and Pendleton, G. N. 1993, ApJ, 413, L101

Link, B., Epstein, R. I., and Priedhorsky, W. C. 1993, ApJ, 408, L81

Mazets, E. P. et al. 1981, Ap \& SS, 80, 3

McBreen, B. \& Metcalfe, L. 1988, Nature, 332, 234

Mitrofanov, I. G. et al., 1992, in Gamma-Ray Bursts, AIP Conference
Proceedings 265, eds: W. S. Paciesas and G. J. Fishman, (New York: AIP),
163

\hangindent=20pt
Meegan, C. A., Fishman, G. J., Wilson, R. B., Paciesas, W. S., Pendleton,
G. N., Horack, J. M., Brock, M. N., and Kouveliotou, C. 1992, Nature, 355,
143

Nemiroff, R. J. 1993, Comm. Astrophys., in press

\hangindent=20pt
Norris, J. P., Cline, T., Desai, U., and Teegarden, B. 1984, Nature, 308, 434

Norris, J. P. et al. 1986, Adv. Space. Res., 6, 19

\hangindent=20pt
Norris, J. P., Davis, S. P., Kouveliotou, C., Fishman, G. J., Meegan, C.
A., Wilson, R. B., and Paciesas, W. S. 1993, in Compton Gamma-Ray
Observatory, AIP Conference Proceedings 280, eds: M. Friedlander, N.
Gehrels, and D. J. Macomb (New York: AIP), 959

\hangindent=20pt
Weisskopf, M. C., Sutherland, P. G., Katz, J. I., and Canizares, C. R.
1978, ApJ, 223, L17

}

\vfill\eject

\cl{\bf FIGURE CAPTIONS }
\hangindent=0pt
\baselineskip=18pt

\bigskip

\noindent {\bf Figure 1:} Ratio between number of time bins higher and
lower than the preceding bin ($\Upsilon$), versus BATSE trigger number. The
bin size is 0.064 seconds for all GRBs, and the energy range used is $\sim$
50 and $\sim$ 300 keV. Most GRBs individually show a time-asymmetry in
their light curves.

\bigskip

\noindent {\bf Figure 2:} Ratio between number of time bins higher and
lower than the preceding bin ($\Upsilon$), versus size of bin, for specific
BATSE trigger numbers and for the complete sample. Bin centers are exactly
factors of two but are slightly staggered to show the one sigma error bars
more clearly.

\bigskip

\noindent {\bf Figure 3:} Light curves for the three GRBs discussed in
text.  Bin size for each plot is 0.064 seconds.  In 3a (BATSE trigger 143),
only the first envelope of pulses is shown.  Note how both individual
pulses and overall burst envelope are time-asymmetric. In 3b (trigger 678)
the envelope is asymmetric while the shorter time scales show no
significant asymmetry, while in 3c (trigger 1606) neither the shorter
time scales nor longer time scales characteristic of the envelope show a
significant amount of asymmetry.

\bigskip

\noindent {\bf Figure 4:} Ratio between number of time bins higher and
lower than the preceding bin ($\Upsilon$), versus size of bin, for the
complete sample of GRBs in the four energy bands of BATSE's Large Area
Detectors.  A significant asymmetry is present for each energy band and for
each time scale.  Bin centers are slightly staggered as in Figure 2.

\vfill\eject

\end